# Impact of Random Loss on TCP Performance in Mobile Ad-hoc Networks (IEEE 802.11): A Simulation-Based Analysis

**Shamimul Qamar**
*Department of Computer Sceience*
CAS, King Saud University
Riyadh,Saudi Arabia

**Kumar Manoj**
*ECED,DPT*
I.I.T Roorkee
India

**Abstract-Initially TCP was designed with the notion in mind that wired networks are generally reliable and any segment loss in a transmission is due to congestion in the network rather than an unreliable medium (The assumptions is that the packet loss caused by damage is much less than 1%) . This notion doesn't hold in wireless parts of the network. Wireless links are highly unreliable and they lose segments all the time due to a number of factors. Very few papers are available which uses TCP for MANET. In this paper, an attempt have been made to justify the use of TCP variants (Tahoe and Reno) for loss of packet due to random noise introduces in the MANET. For the present analysis the simulation has been carried out for TCP variants (Tahoe and Reno) by introduces 0%, 10%, 20% and 30% noise.The comparison of TCP variants is made by running simulation for 0%, 10%, 20% and 30% of data packet loss due to noise in the transmission link and the effect of throughput and congestion window has been examined. During the simulation we have observed that throughput has been decreased when a drop of multiple segments happens, further we have observed in the case of TCP variant (Reno) throughput is better at 1% (Figure 5) which implies a network with short burst of error and low BER, causing only one segment to be lost. When multiple segments are lost due to error prone nature of link, Tahoe perform better than Reno (Figure 13), that gives a significant saving of time (64.28%) in comparison with Reno (Table 4). Several simulations have been run with ns-2 simulator in order to acquire a better understanding of these TCP variants and the way they perform their function. We conclude with a discussion of whether these TCP versions can be used in Mobile Ad hoc Network?**

**Index Term- TCP, Tahoe, Reno, MANET, ns-2, Random noise**

## 1. Introduction

TCP is the widely used for Internet traffic. In the wired network losses are mostly due to congestion. In practice, losses may also be caused from noisy links. This is especially true in case of radio links, e.g. in cellular network, Mobile Ad-hoc network or in satellite links. A link may become in fact completely disconnected for some period of time or it may suffer from occasional interferences (due to shadowing, fading etc.) that cause packets to contain errors and then to be dropped resulting low throughput. Transport protocols should be independent of the technology of the underlying network layers. It should not care whether IP is running over fiber or radio. But in reality, it does matter, since most TCP implementations that have been designed for wired networks, don't cope very well on wireless networks. Different version of TCP in a local network with lossy link are compared and analyzed in [7].Several scheme have been reported in the literature, to alleviate the effect of losses on TCP performances over wireless network (or networks with lossy link) [8]. Very few papers are available which uses TCP for MANET This paper deals with for Tahoe and Reno TCP variants and its variation of error by introducing different level of noise . The ns-2 network simulator used to understand the behaviors and characteristics of TCP variants. Wireless links are highly unreliable and they lose segments all the time due to a number of factors. These include fading, interference, higher bit error rate and mobility related processes such as handovers.

In this paper, we will make a comparison between Tahoe and Reno TCP variants in a noisy environment. Our simulation show that Tahoe copes well in a high segment loss environment. Two algorithms, Slow Start and Congestion Avoidance, modify the performance of TCP's sliding window to solve some problems relating to congestion in the network [4]. The idea of congestion control is for the sender to determine the capacity that is available on the network. The sender, in standard TCP implementation, keeps two state variables for congestion control: a slow-start/congestion window, $cwnd$, and a threshold size, $ssthresh$. These variables are used to switch between the slow start and congestion avoidance algorithms. Slow start provides a way to control initial data flow at the beginning of a communication session and during an error recovery. This is based on received acknowledgements. Congestion avoidance increases the $cwnd$ additively, so that it grows by one segment for each round trip time. These two algorithms are implemented together and work as if they were one. The combined effect of these two algorithms on the *congestion window* is shown in Figure 1. In this paper we report the simulation results of different scenarios. In Section-2, a summary of Tahoe and Reno TCP transport protocols has been reported.

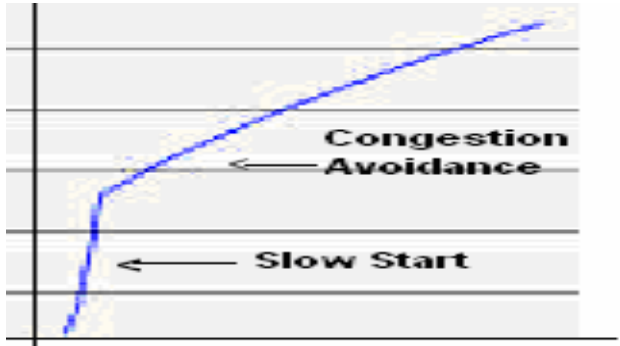

Figure 1. Congestion window

The simulation software and the network simulation setup are described in section 3. The comparison of simulation results is presented in section 4. The conclusion has been reported in section 5.

## 2. TCP Transport Protocol

### 2.1 Tahoe TCP

Modern TCP implementations contain a number of algorithms aimed at controlling network congestion while maintaining good user throughput. Early TCP implementations followed a go-back- model using cumulative positive acknowledgment and requiring a retransmit timer expiration to re-send data lost during transport. These TCPs did little to minimize network congestion. The Tahoe TCP implementation added a number of new algorithms and refinements to earlier implementations. The new algorithms include *Slow-Start*, *Congestion Avoidance*, and *Fast Retransmit* [3]. The refinements include a modification to the round-trip time estimator used to set retransmission timeout values. All modifications have been described elsewhere. The fast retransmit algorithm is of special interest in this paper because it is modified in subsequent versions of TCP. With Fast Retransmit, after receiving a small number of duplicate acknowledgments for the same TCP segment (*dup ACKs*), the data sender infers that a packet has been lost and retransmits the packet without waiting for a retransmission timer to expire, leading to higher channel utilization and connection throughput.

*2.2 Reno TCP* The Reno TCP implementation retained the enhancements incorporated into Tahoe, but modified the Fast Retransmit operation to include *Fast Recovery* [3]. The new algorithm prevents the communication path ("pipe") from going empty after Fast Retransmit, thereby avoiding the need to slow-start to re-fill it after a single packet loss. Fast recovery operates by assuming each dup ACK received represents a single packet having left the pipe. Thus, during Fast recovery the TCP sender is able to make intelligent estimates of the amount of outstanding data. Fast Recovery is entered by a TCP sender after receiving an initial threshold of dup ACKs. This threshold, usually known as *tcprexmtthresh*, is generally set to three. Once the threshold of dup ACKs is received, the sender retransmits one packet and reduces its congestion window by one half. Instead of slow-starting, as is performed by a Tahoe TCP sender, the Reno sender uses additional incoming dup ACKs to clock subsequent outgoing packets. In Reno, the sender's *usable* window becomes where is the receiver's advertised window, is the sender's congestion window, and is maintained at until the number of dup ACKs reaches *tcprexmtthresh*, and thereafter tracks the number of duplicate ACKs. Thus, during Fast Recovery the sender "inflates" its window by the number of dup ACKs it has received, according to the observation that each dup ACK indicates some packet has been removed from the network and is now cached at the receiver. After entering Fast Recovery and retransmitting a single packet, the sender effectively waits until half a window of dup ACKs have been received, and then sends a new packet for each additional dup ACK that is received. Upon receipt of an ACK for new data (called a "recovery ACK"), the sender exits Fast Recovery by setting to Reno's fast recovery algorithm is optimized for the case when a single packet is dropped from a window of data. The Reno sender retransmits at most one dropped packet per round-trip time as shown in Table[1]. Reno significantly improves upon the behavior of Tahoe TCP when a single packet is dropped from a window of data, but can suffer from performance problems when multiple packets are dropped from a window of data. This is illustrated in the Table[2] with three or more dropped packets as shown in Table[3].

Table 1: Round trip times in seconds (at sink nodes)

| Minimal RTT(CN,ON,SPID) | 0.303733(0,4,2) |
|---|---|
| Maximal RTT(CN,ON,SPID) | 0.594773(0,4,49) |
| Average RTT | 0.5565596345 |

Table 2.Average generated & received packet of Tahoe & Reno for different niose.

| Tahoe (at source node) | | | | |
|---|---|---|---|---|
| | 0% | 10% | 20% | 30% |
| Generated packets | 290 | 63 | 61 | 21 |
| Average Packet Size | 538.275 | 561.0084 | 606.0377 | 628.2353 |
| Reno (at sink node) | | | | |
| | 0% | 10% | 20% | 30% |
| Received Packets | 290 | 56 | 45 | 13 |
| Average Packet Size | 538.275 | 531.0714 | 606.0377 | 501.5385 |

The problem is easily constructed in our simulator when a Reno TCP connection with a large congestion window suffers a burst of packet losses after slow-starting in a network with drop-tail gateways (or other gateways that fail to monitor the average queue size).

## 3. Simulation Setup

This section describes four simulation scenarios for 0%, 10%, 20% and 30% of data packet loss due to noise in the transmission link. Each set of scenarios is run for Tahoe, Reno, New-Reno and SACK TCP. The results in this paper are based on simulations using the ns-2 network simulator. The network is setup as per the network diagram in the Figure 2 such as there are 5 nodes from n0 to n5 and node n0 and n2 are connected by bidirectional link of 2 Mbps bandwidth and a propagation delay of 10ms. We have implemented a DropTail queue between node n2 and n3. FTP traffic is sent from node n0 to n4 via shared link (n2-n3) by using TCP transport protocol and CBR traffic is sent from node n1 to n5 using same shared link (n2-n3) by using UDP protocol.

The various simulation parameters used for the present analysis for both (Tahoe and Reno) are like no. of nodes (6), simulation length in seconds(10.47), no. of sending nodes (3), no. of receiving nodes(3), no. of generated packet(593), no. of forwarded packet(1189), number of forwarded bytes(325200), average packet size(548.398).

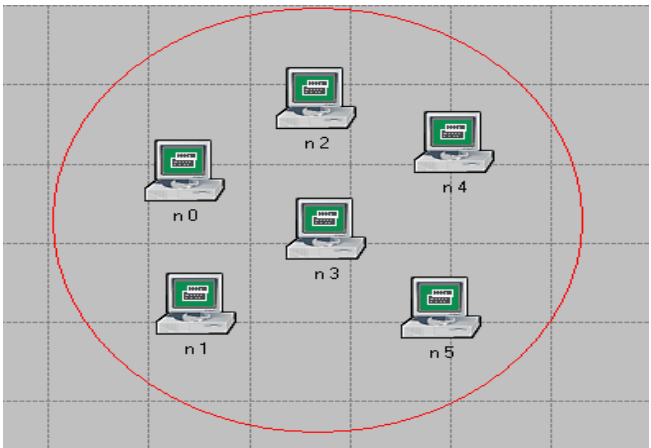

Figure 2. Proposed system network

The simulation has been carried out for 141 simulation seconds in all four scenario viz; (0%, 10%, 20% and 30%) for Tahoe and Reno TCP varients. In the first scenario the network is simulated without noise at the shared link whereas in the second, third and fourth scenario we have introduced a noise of 10%, 20% and 30% are introduced in the shared link (forward link of n2 to n3) respectively.

## 4. Analytical and Simulation Comparisons

In this section we have presented the simulation results for the scenarios mentioned above.

On comparing the throughput of TCP Reno and Tahoe at without loss (0% noise) the throughput remains same as shown in Figure 3-4, whereas on increasing the noise (1%) the throughput of Reno is better as compared with the throughput of Tahoe as shown in Figure 5. Fast Retransmit and Fast Recovery in Reno seem to work well when only one segment is lost. But, as we have seen in the simulation

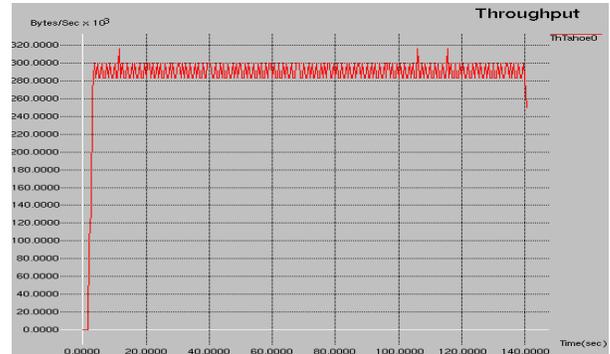

Figure 3. Throughput of TCP Tahoe without loss (0 % noise)

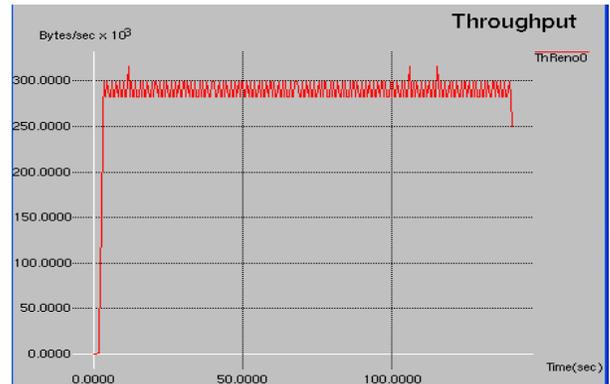

Figure 4. Throughput of TCP Reno without loss (0 % noise)

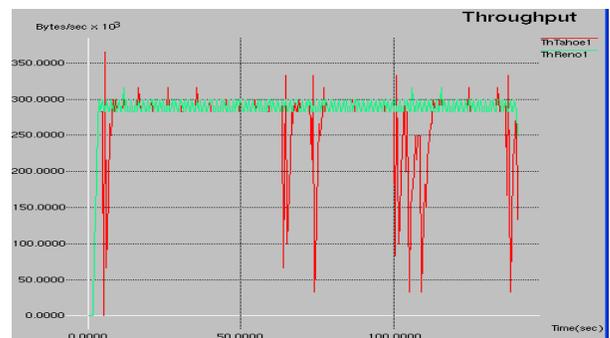

Figure 5. Throughput of TCP Tahoe & Reno with loss (1 % noise)

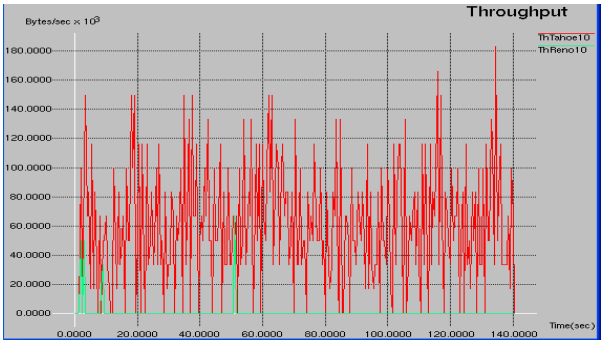

Figure 6. Throughput of TCP Tahoe & Reno with
loss (10 % noise)

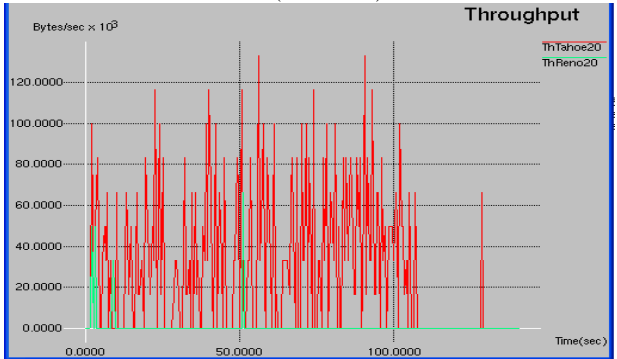

Figure 7. Throughput of TCP Tahoe & Reno with
loss (20 % noise)

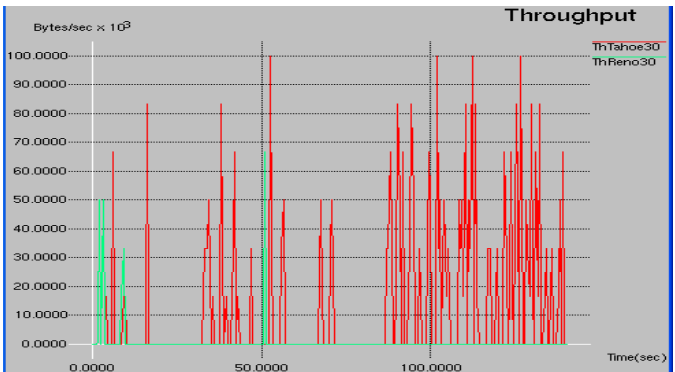

Figure 8. Throughput of TCP Tahoe & Reno with
loss (30 % noise)

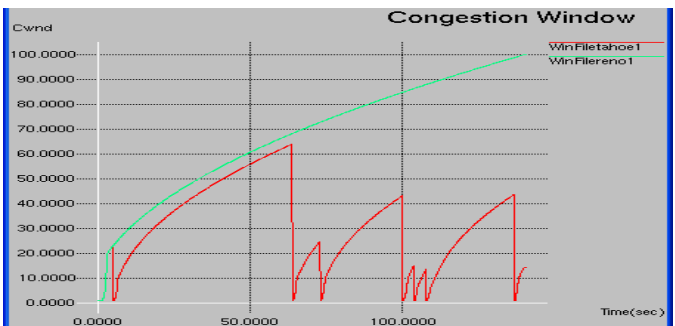

Figure 9. Congestion Window of TCP Tahoe & Reno with loss (1 % noise)

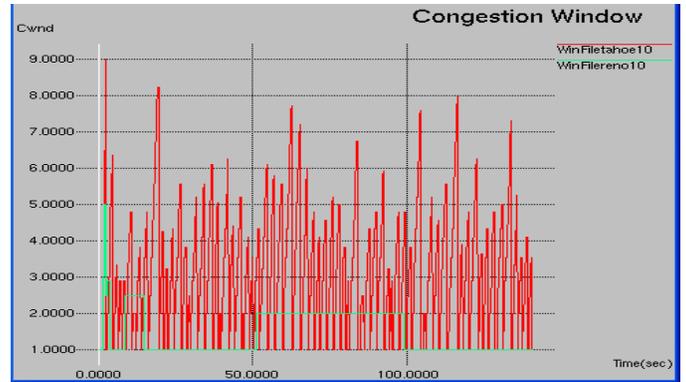

Figure 10. Congestion Window of TCP Tahoe & Reno with loss (10 %
noise)

results, some problems arise when a drop of multiple
segments happens.

As shown in Figures 10-12, at time 4, 15 and 98 sim-secs
when an error causing a loss of multiple segments, a
problem arises with Reno TCP's congestion window
which results in zero throughputs as shown in Figure 6-8.
A loss of multiple segments makes the window close to
half of its size for every dropped segment. Besides, the
usable window is decreased during the fast recovery
phase, as more information is sent; every time a
duplicated acknowledgement is received. When the
congestion window is divided by two for the second time,
the usable window closes to zero, thus blocking the
communication and forcing the retransmission timeout.
Tahoe TCP does not apply Fast Recovery in order to
avoid the problems described previously.

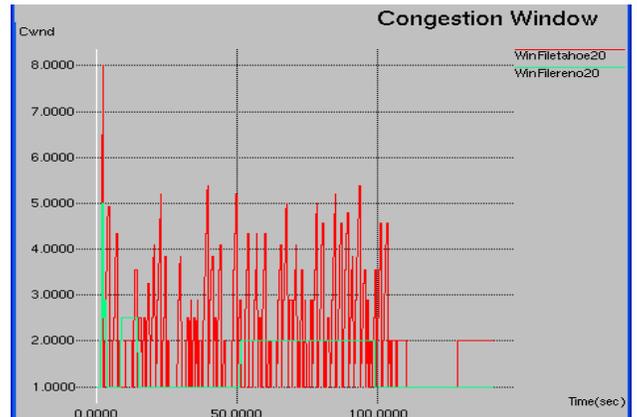

Figure 11. Congestion Window of TCP Tahoe
& Reno with loss (20 % noise)

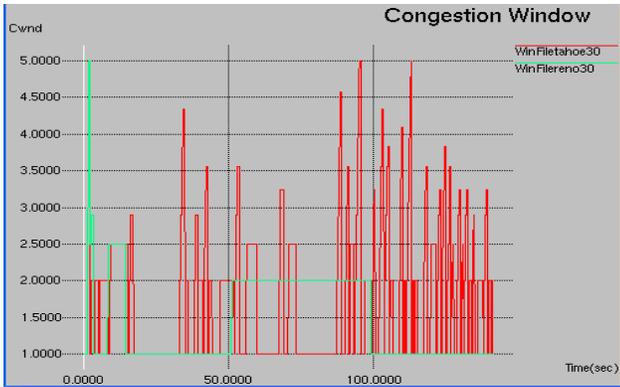

Figure 12. Congestion Window of TCP Tahoe & Reno with loss (30 % noise)

Table 4 End to End delay of Tahoe & Reno variant.

| Tahoe (end2end delay in sec) | |
|---|---|
| Minimal (node,PID) | 0 (2,0) |
| Maximal (node,PID) | 0.244373 (2,49) |
| Average delay | 0.05232697133 |
| Reno (end2end delay in sec) | |
| Minimal delay (CN,ON,PID) | 0.151866(4,0,72) |
| Maximal delay(CN,ON,PID) | 0.254(0,4,23) |
| Average delay | 0.183139041 |

Table 3. Average packet generated and received of Tahoe & Reno variants

| Tahoe (at source node) | | | | |
|---|---|---|---|---|
| | 0% | 10% | 20% | 30% |
| Generated packets | 290 | 76 | 58 | 27 |
| Average Packet Size | 538.275 | 564.4755 | 598.8235 | 630.9091 |
| Tahoe (at sink node) | | | | |
| | 0% | 10% | 20% | 30% |
| Received Packets | 290 | 67 | 44 | 17 |
| Average Packet Size | 538.275 | 532.5373 | 528.6364 | 510.5882 |

When this variant is used, Fast Retransmit is the only algorithm applied. This means that slow start and congestion avoidance will be working when recovering from an error. Tahoe TCP closes the congestion window when the first error is detected. This stops the communication suddenly, but allows the congestion window, and consequently the usable window, to open exponentially as shown in Figure 9. Then the usable window allows the sender to retransmit the lost segments. So, although the usable window is closed, there is always a mechanism that opens it again. The communication is never blocked as in Reno and no waiting for the retransmission timer is needed.

As we discussed above, Tahoe with little modification can be considered in the scenarios where multiple segments are dropped such as in Mobile Ad hoc Networks.

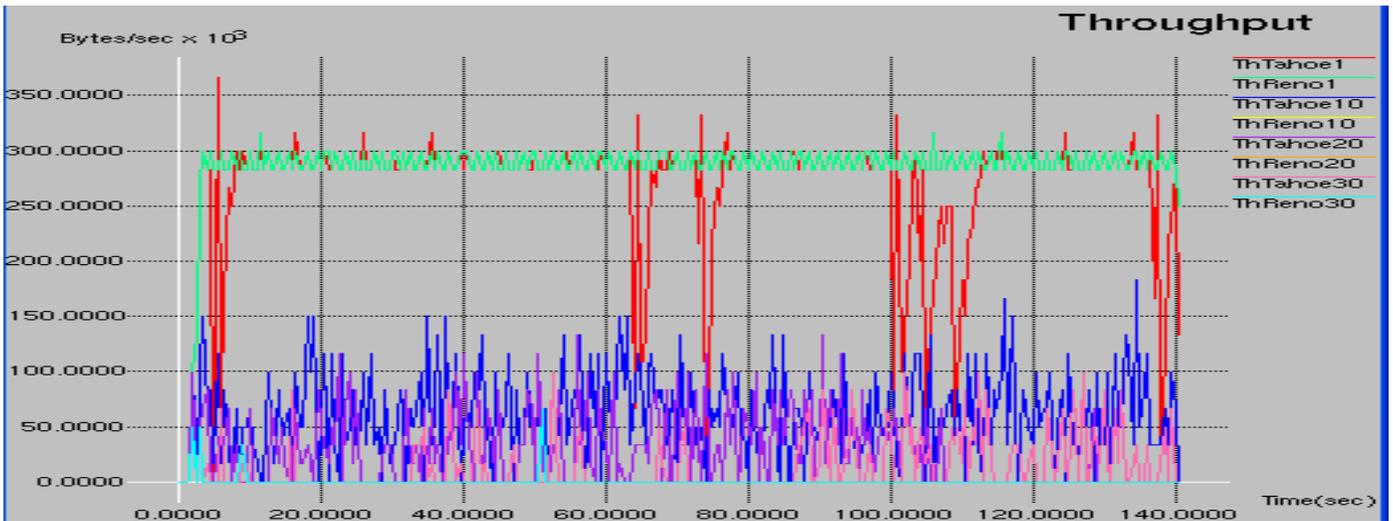

Figure 13. Throughput of TCP Tahoe & Reno with and without loss

## 5. Conclusion

This paper summarizes a study on TCP variants and its congestion control mechanism. We have analyzed from the simulated results in that Tahoe performs better than Reno in multiple data packets loss that gives a significant saving of time (64.28%) in comparison with Reno as shown in (Table 4). The comparison of TCP variants is made by running simulation for 0%, 10%, 20% and 30% of data packet loss due to noise in the transmission link and the effect of throughput and congestion window has been examined in Figure 13.

Thus, TCP Reno implementation could be useful when TCP is working in a reliable network with short bursts of errors and low BER, causing only one segment to be lost. Tahoe can be considered as transport protocol for Mobile Ad hoc network where multiple segments are dropped due to error prone nature of radio link Table [2-3]. Our future work is to focus on the implementation of modified Tahoe in MANET.

## Authors:

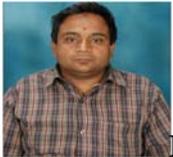

**Dr. Shamimul Qamar**, Professor, Computer Science and Information System department, CAS, King Saud University, Riyadh, has been recognized as an eminent scholar in the field of Compuer Engineering. He has done his B.Tech from MMMEC Gorakhpur, M.Tech from AMU, Aligarh and earned his Ph.D. degree from IIT Roorkee, India with highly honorable grade. Prof.  Qamar has a wide teaching experience in various Engineering colleges. He has research interests in Image processing, Internet Applications, Multimedia systems, computer network and software engineering. He has published several research papers in reputed national/international Journals and conference. He is also a technical programme committee member in international mobility conference, Singapore. He is a life time member of international association of Engineers and a life member of Indian Society of technical educational. His technical depth and interest resulted in setting up a research lab according to latest technical innovations.

**Kumar Manoj** (kumardpt@iitr.ernet.in), member of IEEE, ACEEE, IAENG, ISOC, NSBE (USA) received B.Sc. (Engg.), M.Tech. (Electronics). He has published over Fifty nine research papers in national and International journals/conferences and supervised more than 40 projects/ dissertation of M.Tech. & B.Tech. students. He started his career as R & D Engineer in various MNC companies in the field of Power Electronics then joined teaching profession as a Asstt. Prof. in Graphic Era University. He is a visiting faculty of various Govt. Engg. College. His many research papers have been awarded by National and International Committees/Conference. Presently he is pursuing research work at IIT Roorkee, India in the field of Wireless Communication under Ministry of HRD, Government of India fellowship.  His research interests include the design and control of personal communication networks, mobile multicasting, protocol design and implementation for a mobile integrated services wireless radio network, and high-speed networking.

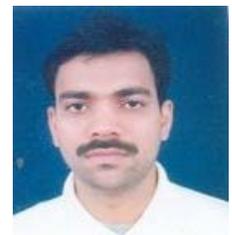